\documentclass[titlepage,prl,byrevtex,amsmath,tightenlines,preprint,
nofootinbib]{revtex4}

\usepackage{graphicx}
\usepackage{bm}
\newcommand{\beq}{\begin{equation}}
\newcommand{\eeq}{\end{equation}}
\newcommand{\eq}[1]{Eq.~(\ref{#1})}

\begin{document}

\preprint{UK/09-05}

\title {Three-Loop Radiative-Recoil Corrections to Hyperfine Splitting in Muonium: Diagrams with Polarization Loops}
\author {Michael I. Eides}
\altaffiliation[Also at ]{Petersburg Nuclear Physics Institute,
Gatchina, St.Petersburg 188300, Russia}
\email{eides@pa.uky.edu, eides@thd.pnpi.spb.ru}
\affiliation{Department of Physics and Astronomy,
University of Kentucky, Lexington, KY 40506, USA,}
\author{Valery A. Shelyuto}
\email{shelyuto@vniim.ru}
\affiliation{D. I.  Mendeleev Institute of Metrology,
St.Petersburg 190005, Russia}

\begin{abstract}

We consider three-loop radiative-recoil corrections to hyperfine splitting in muonium generated by the diagrams with electron and muon vacuum polarizations.
We calculate single-logarithmic and nonlogarithmic contributions of order $\alpha^3(m/M)E_F$  generated by gauge invariant sets of diagrams with electron and muon polarization insertions in the electron and muon factors. Combining these corrections with the older results we obtain total contribution to hyperfine splitting generated by all diagrams with electron and muon polarization loops. Calculation of this contribution completes an important stage in the implementation of the program of reduction of the theoretical uncertainty of hyperfine splitting below 10 Hz. The new results improve the theory of hyperfine splitting, and affect the value of the electron-muon mass ratio extracted from experimental data on the muonium hyperfine splitting.

\end{abstract}


\maketitle
Muonium is a purely electrodynamic bound state, and existence of highly accurate experimental results \cite{lbdd} makes it the best system for comparison of the precise quantum electrodynamic theory  of hyperfine splitting with experiment \footnote{See also recent discussion \cite{bl2009} of experimental feasibility of precise spectroscopy with another purely electrodynamic bound system, dimuonium.}. The hyperfine splitting interval is proportional to the electron-muon mass ratio, and the current theoretical prediction is
\beq \label{hypthval}
\Delta E^{th}_{HFS}(Mu)=4~463~302~904~(518)~(30)~(70)~\mbox{Hz},
\eeq
\noindent
where the first error is due to the experimental error of direct measurement of electron-muon mass ratio $m/M$, the second error is due to the experimental uncertainty of the fine structure constant $\alpha$, and the third error is an estimate of yet uncalculated theoretical corrections (for more details see \cite{egsbook} and \cite{mt2008}). The uncertainty  of the electron-muon mass ratio dominates in the balance of errors, and therefore measurement of hyperfine splitting is the best source for the precise value of this mass ratio. We see from \eq{hypthval} that calculation of all theoretical corrections with magnitude above $10$ Hz would improve accuracy of the electron-muon mass ratio. There are three series of yet uncalculated corrections of such magnitude \cite{egsbook}: a) single-logarithmic and nonlogarithmic radiative-recoil corrections  of order $\alpha^3(m/M)E_F$, b) nonlogarithmic contributions of order $(Z\alpha)^3(m/M)E_F$, and c) nonlogarithmic contributions of order $\alpha(Z\alpha)^2(m/M)E_F$ ($Z$ is the nucleus charge, $Z=1$ for muonium). In this paper we complete the calculation of all single-logarithmic and nonlogarithmic radiative-recoil corrections connected with the electron and muon polarization loops.

\begin{figure}[htb]
\includegraphics[height=1.8cm]{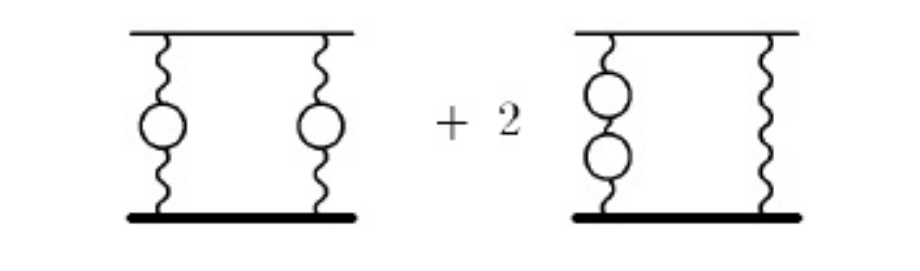}
\caption{\label{onelooppolrecfhsfig}
Graphs with two one-loop polarization insertions}
\end{figure}

\begin{figure}[htb]
\includegraphics[height=2.cm]{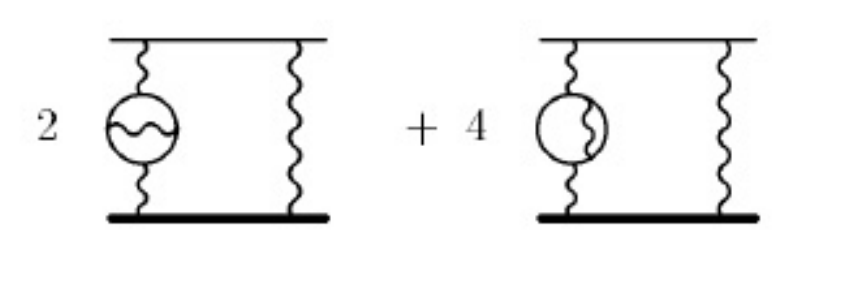}
\caption{\label{twolooppolrecfhsfig}
Graphs with two-loop polarization insertions}
\end{figure}

\begin{figure}[htb]
\includegraphics[height=2.cm
]{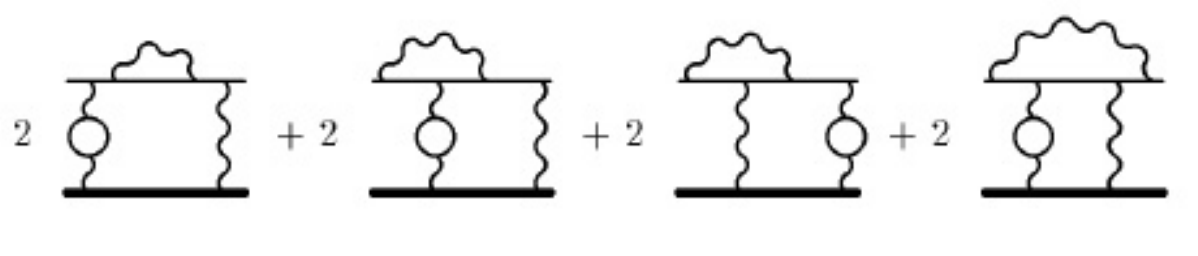}
\caption{\label{ellineinsrechfsfig}Graphs with radiative photon insertions}
\end{figure}

\begin{figure}[htb]
\includegraphics[height=2.5cm]{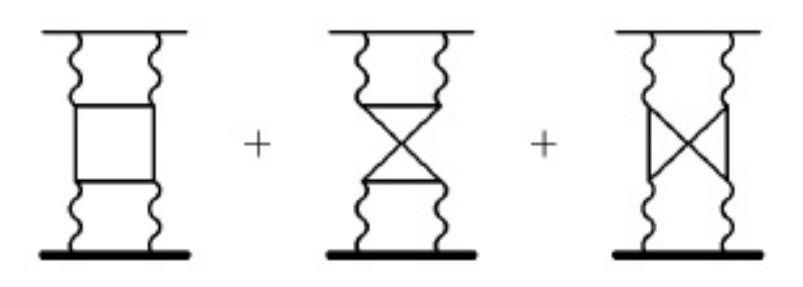}
\caption{\label{lihlightrechfsfig} Graphs with light by light
scattering insertions}
\end{figure}

\begin{figure}[htb]
\includegraphics[height=2.2cm]{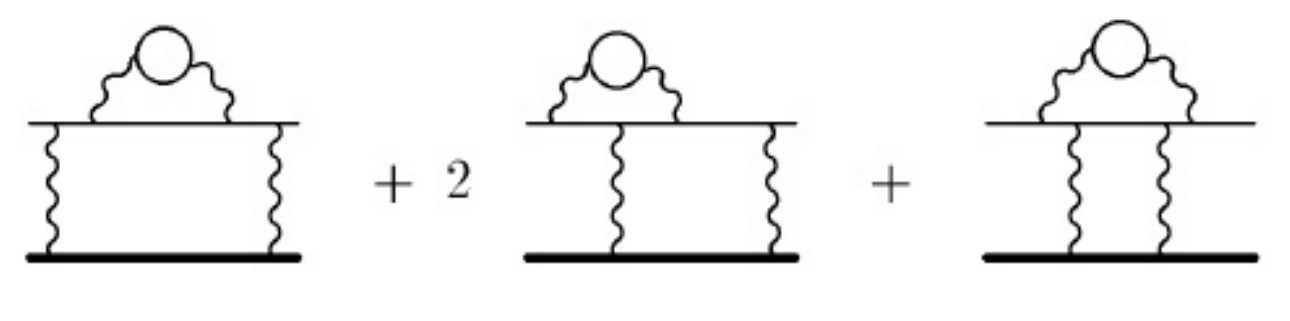}
\caption{\label{polellineinsrechfsfig} Graphs with polarization
insertions in the electron factor}
\end{figure}

\begin{figure}[tbh]
\center\includegraphics[height=2.2cm]{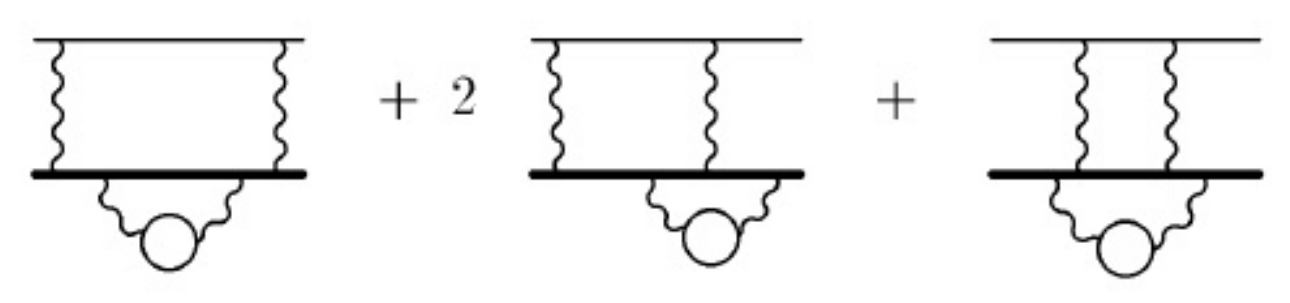}
\caption{\label{elpol} Graphs with polarization
insertions in the muon factor}
\end{figure}

The radiative-recoil corrections of order $\alpha^3(m/M)E_F$ are enhanced by large logarithm of the muon-electron  mass ratio $M/m$ \cite{es0}. The leading logarithm cubed and logarithm squared contributions are generated by the graphs with the electron closed loops in Figs.\ \ref{onelooppolrecfhsfig}-\ref{lihlightrechfsfig} (and by the diagrams with the crossed exchanged photon lines), and were calculated long time ago \cite{es0,eks89}. Single-logarithmic and nonlogarithmic terms of order
$\alpha^3(m/M)E_F$ are generated by all diagrams in
Figs.~\ref{onelooppolrecfhsfig}-\ref{lihlightrechfsfig}, by the respective
graphs with the muon loops, by the graphs with polarization and radiative photon insertions in the muon line, and also by the three-loop graphs with  radiative photons in the electron and/or muon lines but without polarization loops.  Below we calculate three-loop single-logarithmic and nonlogarithmic  radiative-recoil corrections generated by the diagrams in Figs.~\ref{polellineinsrechfsfig}, \ref{elpol} both with the electron and muon loops. We combine these corrections with the earlier results, and present complete results for all radiative-recoil corrections generated by the diagrams with electron and muon polarizations.

There are numerous technical problems connected with calculation of the diagrams in Figs.~\ref{polellineinsrechfsfig}, \ref{elpol}. As usual individual diagrams are ultraviolet and infrared divergent. In addition, the integrands in the Feynman parametrization for the exchanged integrals do not admit expansion in the small mass ratio before integration generating spurious divergences. This problem is solved by direct integration over exchanged momenta in four-dimensional Euclidean space. After angular integrations we used the method of overlapping small and large momentum expansions of the integrand to obtain systematic series over powers of the small electron-muon mass ratio and its logarithm \cite{yaf88,eksann2}. Ultraviolet divergences are really not a problem and were dealt with by usual subtractions.

\begin{figure}[htb]
\includegraphics[height=2.2cm]{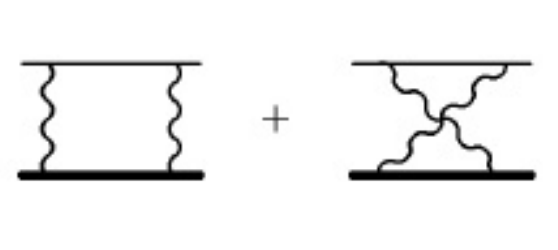}
\caption{\label{twophothfsfig}Diagrams with two-photon exchanges}
\end{figure}

By far the largest challenge is connected with the spurious infrared divergences. Gauge invariant sets  of diagrams in Figs.~\ref{polellineinsrechfsfig}, \ref{elpol} are infrared finite, but each individual diagram is infrared divergent even in the infrared soft Yennie gauge. In order to obtain accurate results we need to separate the would be infrared divergent terms in the integrands, and explicitly cancel them analytically in the sum of contributions of different diagrams. In order to explain how such cancelation is achieved we start with the skeleton diagrams with two-photon exchanges in Fig.~\ref{twophothfsfig}. All three-loop diagrams in Figs.~\ref{polellineinsrechfsfig}, \ref{elpol} may be interpreted as radiative corrections to the skeleton diagrams. We consider the momentum integrand in the skeleton case as a product of the skeleton electron (muon) factor $L_{\mu\nu}(k)$  and the remaining part of the diagram. The factor $L_{\mu\nu}(k)$ is the Compton scattering amplitude for virtual photons. Contributions to HFS generated by the diagrams in Figs.~\ref{polellineinsrechfsfig}, \ref{elpol} are described by the integrals similar to the integrals for the diagrams in Fig.~\ref{twophothfsfig}. The only difference is that for the diagrams in Figs.~\ref{polellineinsrechfsfig}, \ref{elpol} we include in the integral radiatively corrected virtual Compton scattering amplitude instead of the skeleton one. A generalized low energy theorem holds for the virtual Compton scattering amplitude with subtracted anomalous magnetic moment contribution (see more on this subtraction below) \cite{yaf88,eksann2}. According to this theorem the electron (muon) factor $L_{\mu\nu}(k)$ is suppressed by an additional factor $k^2/m^2$ ($k^2/M^2$) in comparison with the respective skeleton factor. Due to this suppression the integrals corresponding to the diagrams in Figs.~\ref{polellineinsrechfsfig}, \ref{elpol} are infrared finite, and the integration region with momenta less than the electron mass are additionally suppressed. As  a result, we can omit the atomic scale external virtualities of order $m\alpha$,  and calculate matrix elements in the scattering regime between the free electron and muon spinors. We use the Feynman gauge to obtain matrix elements of the gauge invariant sets of diagrams in Figs.~\ref{polellineinsrechfsfig}, \ref{elpol}. We have derived some useful identities for the integrands that allowed to cancel the would be infrared divergent terms in the integrands before integration. Insertions of polarization operators in the diagrams in Figs.~\ref{polellineinsrechfsfig}, \ref{elpol} is taken care of by using the massive photon propagator for radiative photons (but not for exchanged photons) with the photon mass squared $\lambda^2=4m^2/(1-v^2)$ or $\lambda^2=4M^2/(1-v^2)$ for the the electron and muon polarization loops, respectively. These massive propagators require an additional integration over velocity $v$ with the weight $\int_0^1 dvv^2(1-v^2/3)/(1-v^2)$.

The diagrams in Fig.~\ref{polellineinsrechfsfig} with the electron polarization loops generate nonrecoil and logarithm squared, single-logarithmic, and nonlogarithmic radiative-recoil contributions to HFS. It turns out that the gauge invariant anomalous magnetic moment in these diagrams does not generate radiative-recoil corrections (see, e.g., \cite{jetp94,eksann2}). Then the radiatively corrected electron factor $L_{\mu\nu}$ provides an additional suppression factor $k^2/m^2$ in the skeleton integral over the exchanged momenta. The wide integration region between the electron and muon masses $m\leq k\leq M$ remains logarithmic. We calculated the nonrecoil contribution numerically, the logarithm squared and single-logarithmic terms analytically, and the nonlogarithmic term numerically. The logarithm squared terms is already well known \cite{eks89},  and the single-logarithmic and nonlogarithmic contributions are as follows

\begin{equation} \label{eline1}
\Delta E=
\Biggl[\biggl(\pi^2 -\frac{53}{6}~\biggr)\ln{\frac{M}{m}} +
7.0807\Biggr]\frac{\alpha^3}{\pi^3}\frac{m}{M}E_F.
\end{equation}

The electron factor with muon polarization insertions in the diagrams in Fig.~\ref{polellineinsrechfsfig} provides an additional suppression factor $k^2/M^2$, and  lifts characteristic integration momenta to the scale of the muon mass. Then these  diagrams do not generate nonrecoil and logarithmic contributions to HFS. The respective leading recoil correction is a pure number, that we calculated it numerically

\begin{equation} \label{eline2}
\Delta E=-1.3042~\frac{\alpha^3}{\pi^3}\frac{m}{M}E_F.
\end{equation}

Consider now diagrams in Fig.~\ref{elpol} with radiative corrections in the muon line. The radiatively corrected muon factor provides an additional  suppression factor $k^2/M^2$ and the integral over exchanged momenta is nonlogarithmic.                                       There is, however, another source of large logarithms in the case of electron polarization insertions in Fig.~\ref{elpol}. The electron polarization insertions in the diagrams in  Fig.~\ref{elpol} enter in the asymptotic regime, and the respective leading contribution to HFS is the product of the leading asymptotic term in the high momentum expansion of the electron polarization operator and the radiative-recoil correction to HFS generated by the one-loop muon factor without polarization insertions. In other words the leading logarithmic contribution to HFS generated by the diagrams in Fig.~\ref{elpol} is obtained from the respective nonlogarithmic contribution of the diagrams without polarization insertions by substitution of the running coupling constant $\alpha(M)$ for radiative photons. We calculated also the nonlogarithmic contribution and obtained

\beq
\Delta E=\biggl[
\biggl(3\zeta (3) - 2\pi^2\ln{2}+\frac{13}{4}\biggr)\ln{\frac{M}{m}}
+ 12.227(2)\biggr]
\frac{\alpha^3}{\pi^3}\frac{m}{M}E_F.
\label{muline1}
\eeq

The muon factor with muon polarization loops in the diagrams in Fig.~\ref{elpol} again provides the suppression factor $k^2/M^2$, but no enhancements. As a result these diagrams generate only nonlogarithmic radiative-recoil corrections. After numerical calculations we obtained

\beq \label{muline2}
\Delta E=-0.931\frac{\alpha^3}{\pi^3}\frac{m}{M}E_F.
\eeq

Other single-logarithmic and nonlogarithmic three-loop radiative-recoil corrections generated by the diagrams with electron and muon  polarization insertions were obtained earlier, and we collect respective results below. Single-logarithmic and nonlogarithmic radiative-recoil corrections generated by the diagrams with two electron or two muon loops in Fig.~\ref{onelooppolrecfhsfig} are \cite{egs01}

\beq \label{finemuoneloop}
\Delta E
=\biggl[- \biggl(\frac{2\pi^2}{3} + \frac{25}{9} \biggr)\ln{\frac{M}{m}}
- \frac{4\pi^2}{9} - \frac{535}{108}\biggr]
\frac{\alpha^3}{\pi^3}\frac{m}{M}E_F.
\eeq

\begin{figure}[htb]
\includegraphics[height=2.cm]{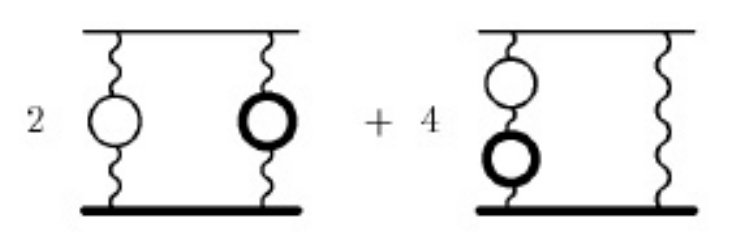}
\caption{\label{mixedloops}Graphs with both the electron and muon (bold) loops}
\end{figure}

The diagrams with one electron and one muon loop in Fig.~\ref{mixedloops} produce only single-logarithmic and nonlogarithmic contributions to HFS \cite{egs01}

\beq   \label{finmixoneloop}
\Delta E=
\biggl[
\biggl(\frac{2\pi^2}{3} - \frac{20}{9} \biggr) \ln{\frac{M}{m}}
+ \frac{\pi^2}{3} - \frac{53}{9} \biggr]\frac{\alpha^3}{\pi^3}
\frac{m}{M}E_F.
\eeq

The single-logarithmic and nonlogarithmic radiative-recoil contributions to HFS generated by the diagrams in Fig.~\ref{twolooppolrecfhsfig} with
two-loop electron or two-loop muon polarization insertions were calculated analytically \cite{egs01}

\begin{eqnarray}
&&\Delta E =\biggl[
- \biggl(6\zeta (3) + \frac{13}{4}\biggr)\ln{\frac{M}{m}}
-\frac{97}{8}\zeta{(3)} - 16\mbox{Li}_4\biggl(\frac{1}{2}\biggr)
\nonumber\\*
&&+ \frac{2\pi^2}{3}\ln^2{2} - \frac{2}{3}\ln^4{2}
+ \frac{5\pi^4}{36} - \frac{\pi^2}{4} + \frac{7}{16}\biggr]
\frac{\alpha^3}{\pi^3}
\frac{m}{M}E_F.
\label{fintwoloop}
\end{eqnarray}

The single-logarithmic and nonlogarithmic radiative-recoil contributions generated by the diagrams with electron or muon polarization insertions in the exchanged photons in Fig.~\ref{ellineinsrechfsfig} have the form \cite{egs03}

\beq \label{eeresult}
\Delta E=
\left(\frac{10}{3}\ln{\frac{M}{m}}+
8.6945\right)\frac{\alpha^3}{\pi^3}
\frac{m}{M}E_F.
\eeq

\begin{figure}[htb]
\includegraphics[height=2.cm
]{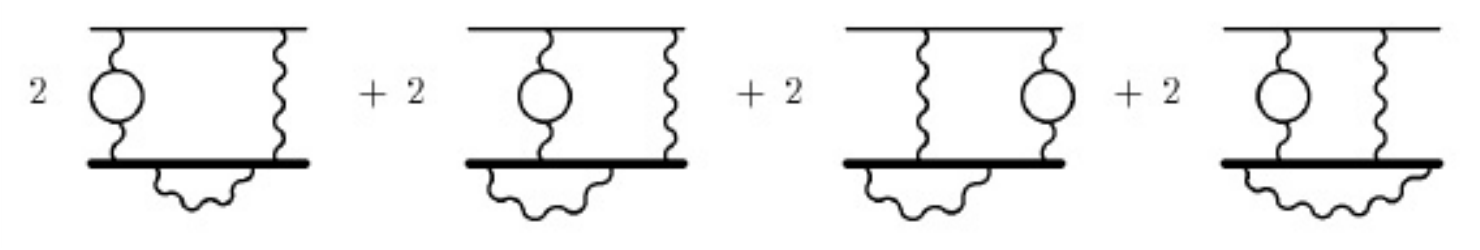}
\caption{\label{em} Muon line and electron vacuum polarization}
\end{figure}

The diagrams in Fig.~\ref{em} with the electron or muon polarization insertions in the exchanged photons generate only single-logarithmic and nonlogarithmic radiative-recoil contributions to HFS \cite{egs03}

\beq
\Delta  E=\biggl[\biggl(6\zeta{(3)} - 4 \pi^2 \ln{2} + \frac{5}{2}
\biggr) \ln{\frac{M}{m}}
+ 23.8527\biggr]
\frac{\alpha^3}{\pi^3} \frac{m}{M} E_F.
\label{mueresult}
\eeq

Combining all three-loop single-logarithmic and nonlogarithmic radiative-recoil corrections to hyperfine splitting due to electron and muon polarization loops in \eq{eline1}) - \eq{mueresult} we obtain

\beq
\Delta E_{tot}=\biggl[\biggl(3\zeta(3)- 6 \pi^2\ln{2}+\pi^2-8\biggr) \ln{\frac{M}{m}}
+27.666~(2)\biggr]\frac{\alpha^3}{\pi^3}\frac{m}{M}E_F.
\label{finrespol}
\eeq

\begin{figure}[htb]
\includegraphics[height=2.2cm]{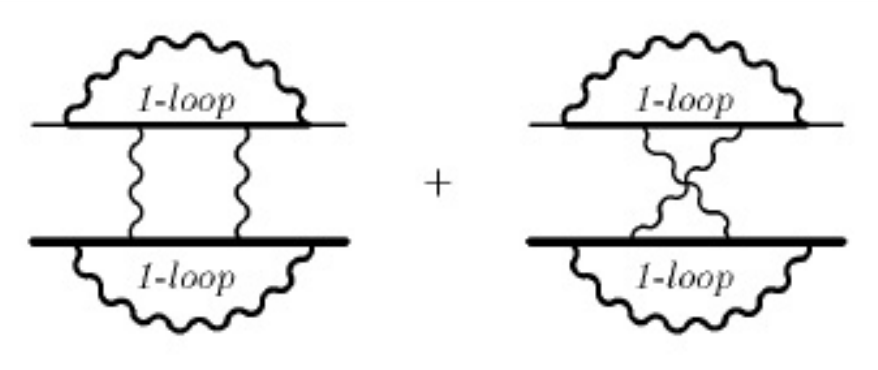}
\caption{\label{radrecdiag} Diagrams with two fermion factors}
\end{figure}

For completeness let us mention that the three-loop radiative-recoil correction generated by the diagrams with one-loop fermion factors (and without polarization loops) in Fig.~\ref{radrecdiag}  is also known \cite{egs04}

\beq \label{newrsult}
\Delta E =\biggl(-\frac{15}{8}\zeta{(3)} + \frac{15\pi^2}{4}
\ln{2} + \frac{27\pi^2}{16} - \frac{147}{32} \biggr)
\frac{\alpha^3}{\pi^3}\frac{m}{M}E_F.
\eeq

The only still unknown single-logarithmic and nonlogarithmic three-loop radiative-recoil corrections are generated by the gauge invariant sets of diagrams with two-loop fermion factors without polarization insertions, and the diagrams with light-by-light insertions in the exchanged photons. Calculation of these corrections is a task for the future.

The total result for all known three-loop single-logarithmic and nonlogarithmic radiative-recoil corrections to hyperfine splitting is given by the sum of the  contributions in \eq{finrespol} and \eq{newrsult}

\beq
\Delta E_{tot}=\biggl[\biggl(3\zeta(3)- 6 \pi^2\ln{2}+\pi^2-8\biggr) \ln{\frac{M}{m}}
+63.127~(2)\biggr]\frac{\alpha^3}{\pi^3}\frac{m}{M}E_F.
\label{finres}
\eeq

\noindent
Numerically this contribution to HFS in muonium is

\begin{equation}
\Delta E_{tot}=-34.7~\mbox{Hz}.
\end{equation}

As was explained above the current goal in the theory of hyperfine splitting is to reduce the theoretical uncertainty below 10 Hz (see a more detailed discussion in \cite{egs01r,egsbook,egs03}). The muon and electron polarization operator contributions and other corrections collected in \eq{finres}, together with the results of comparable magnitude in \cite{egs98,my,rh,ks2001,cek2002,ksv2008} constitute a next step toward achievement of this goal. Phenomenologically, the improved accuracy of the theory of hyperfine splitting would  lead to a reduction of the uncertainty of the value of the electron-muon mass ratio derived from the experimental data \cite{lbdd} on hyperfine splitting (see, e.g., reviews in \cite{egs01r,egsbook,mt2008}).

This work was supported by the NSF grant PHY--0757928. V.A.S. was also
supported in part by the RFBR grants 06-02-16156 and 08-02-13516, and by the DFG grant GZ  436 RUS 113/769/0-3.

\end{document}